\documentclass[prd,aps,secnumarabic,showpacs]{revtex4}
\usepackage{amsmath}
\begin{document}
\title{Neutrino mixing with broken $S_3$ symmetry}
\renewcommand{\thefootnote}{\fnsymbol{footnote}}
\author{Duane A. Dicus$^{1,}$\footnote{Electronic address: dicus@physics.utexas.edu}, 
Shao-Feng Ge$^{1,2,}$\footnote{Electronic address: gesf02@mails.tsinghua.edu.cn}, 
and Wayne W. Repko$^{3,}$\footnote{Electronic address: repko@pa.msu.edu}}
\affiliation{$^1$Physics Department, University of Texas, Austin, TX 78712  \\
$^2$Center for High Energy Physics, Tsinghua University, Beijing 100084, China  \\
$^3$Department of Physics and Astronomy, Michigan State University, East Lansing MI 48824}
\renewcommand{\thefootnote}{\arabic{footnote}}

\date{\today}

\begin{abstract}
We explore the consequences of assuming that the neutrino mass matrix is a linear combination of the matrices of a three dimensional representation of the group $S_3$ and that it has one zero mass eigenvalue. When implemented, these two assumptions allow us to express the transformation matrix relating the mass eigenstates to the flavor eigenstates in terms of a single parameter which we fit to the available data. 
\end{abstract}
\maketitle

\section{Introduction}
Following the discovery of neutrino oscillations, there has been considerable progress in determining values for the neutrino mass differences $m_i^2-m_j^2$ and for the mixing angles relating the mass eigenstates to the flavor eigenstates. The most recent fits suggest that one of the mixing angles is approximately zero and another has a value that implies a mass eigenstate that is nearly an equal mixture of $\nu_\mu$ and $\nu_\tau$. If these conclusions were exact, then they could be accommodated by postulating a neutrino mass matrix having a symmetry based on a three dimensional representation of the permutation group $S_3$. This connection has been extensively studied in the papers listed in Ref.\,\cite{intro}. The approach taken here is to retain a remnant of the $S_3$ symmetry, assume that one neutrino mass is zero, and see what this implies about the final form of the neutrino mass matrix and transformation between mass and flavor eigenstates. 

For Majorana neutrinos the most general form of the mass matrix is
\begin{equation}\label{M}
M_{\nu}\,=\,\left(\begin{array}{ccc}
                 A & B_1 & B_2  \\
                 B_1 & C_1 & D  \\
                 B_2 & D & C_2  \end{array} \right)
\end{equation}
Experiment seems to show approximate $\mu-\tau$ symmetry in the sense that one mass eigenstate has an almost equal probability of being $\nu_{\mu}$ or $\nu_{\tau}$.  To realize this with $M_{\nu}$ requires $B_1\approx\,B_2$ and $C_1\approx\,C_2$. As mentioned above, exact $\mu-\tau$ symmetry can be nicely modelled using a 3-dimensional representation of the finite group $S_3$. However, suppose $\mu-\tau$ symmetry is not exact but we assume $M_{\nu}$ can still be expressed by the matrices of $S_3$. This ansatz, together with the assumption that one of the neutrino mass eigenvalues is zero, as required, for example, by the minimal seesaw model, allows us to derive two relations among the mixing angles and to predict all of the mixing angles in terms of one parameter.

In the next section we review the conditions imposed by $S_3$ on the elements of $M_{\nu}$. In Sec. 3 we discuss the effect of these conditions on the minimal seesaw model. Following that, in Sec. 4, we find the eigenvalues and eigenstates when one mass eigenvalue is zero. Then, in Sec. 5, we write $A,\ldots,D$ of $M_{\nu}$ in terms of the mixing angles in the usual way and use the conditions derived from $S_3$ and from having one eigenvalue zero to find relations among the mixing angles. We are able to express these angles in terms of one parameter. In the last section we summarize our conditions on the mixing angles and compare our predictions with experiment. Finally, in an Appendix we discuss why it is possible to study neutrino mixing separately from the charged lepton sector.

\section{Conditions on the mass matrix from $S_3$}
The three dimensional representation of $S_3$ is well known.  
Nevertheless, for clarity, we will repeat it here.  Each line of the following gives
the elements that belong to a particular class
\begin{eqnarray}
D(e)\,&=&\,\left(\begin{array}{ccc}
              1 & 0 & 0  \\
              0 & 1 & 0  \\
              0 & 0 & 1 \end{array} \right)   \\
D(a)\,&=&\,\left(\begin{array}{ccc}
              0 & 0 & 1   \\
              1 & 0 & 0   \\
              0 & 1 & 0  \end{array} \right)\,,\,\,\,\,D(b)\,=\,\left(\begin{array}{ccc}
                                                                     0 & 1 & 0  \\
                                                                     0 & 0 & 1  \\
                                                                     1 & 0 & 0  \end{array} \right)  \\
D(c)\,&=&\,\left(\begin{array}{ccc}
              1 & 0 & 0  \\
              0 & 0 & 1  \\
              0 & 1 & 0  \end{array}\right)\,,\,\,\,D(d)\,=\,\left(\begin{array}{ccc}
                                                                   0 & 0 & 1  \\
                                                                   0 & 1 & 0  \\
                                                                   1 & 0 & 0  \end{array} \right)\,,\,\,
D(f)\,=\,\left(\begin{array}{ccc}
              0 & 1 & 0  \\
              1 & 0 & 0  \\
              0 & 0 & 1  \end{array} \right)
\end{eqnarray}
This is a reducible representation and the sum of the elements in each class commutes with every element of the group.
If we define
\begin{equation}\label{classes}
D_1\,=\,D(e)\,,\,\,\,\,D_2\,=\,D(a)+D(b)\,,\,\,\,D_3\,=\,D(c)+D(d)+D(f)
\end{equation}
then the most general mass matrix, invariant under $S_3$, is
\begin{equation}\label{Minvar}
M\,=\,\alpha\,D_1+\beta\,D_2+\gamma\,D_3\,\equiv\,\left(\begin{array}{ccc}
                                                          A & B & B  \\
                                                          B & A & B  \\
                                                          B & B & A  \end{array} \right)
\end{equation}
Clearly this is not general enough but we can get a matrix which still respects $\mu-\tau$ symmetry by breaking $S_3$ with $D(c)$,
\begin{equation}\label{S3c}
M\,=\,\alpha\,D_1+\beta\,D_2+\gamma\,D(c)\,=\,\left(\begin{array}{ccc}
                                                \alpha+\gamma & \beta & \beta  \\
                                                  \beta & \alpha & \beta+\gamma  \\
                                                 \beta & \beta+\gamma & \alpha  \end{array} \right)\,.
\end{equation}
This has the additional condition $A+B=C+D$ necessary for tri-bimaximal mixing.

But suppose $\mu-\tau$ symmetry is not exact so we break $S_3$ in all possible ways
\begin{equation}\label{S3cdf}
M\,=\,\alpha\,D_1+\beta\,D_2+\gamma\,D(c)+\delta\,D(d)+\epsilon\,D(f)\,=\,\left(\begin{array}{ccc}
                                                                      \alpha+\gamma & \beta+\epsilon & \beta+\delta  \\
                                                                      \beta+\epsilon & \alpha+\delta & \beta+\gamma  \\
                                                                      \beta+\delta & \beta+\gamma & \alpha+\epsilon \end{array} \right)
\end{equation}
where we don't include $D(a)$ or $D(b)$ because we would have to add them to get a symmetric matrix and their
sum is $D_2$, and we omit $D_3$ because it just adds the same amount to each matrix element.
Thus we get something of the form of Eq.\,(\ref{M}) but the important thing is that however we break the $\mu-\tau$
symmetry there are still two relations among the elements,
\begin{eqnarray}
2A+B_1+B_2\,&=&\,C_1+C_2+2D  \label{S1}  \\
B_1-B_2\,&=&\,C_2-C_1\,.  \label{S2}
\end{eqnarray}
These remnants of the $S_3$ symmetry are what we will use to restrict 
the parameters of the minimal seesaw model and to restrict
the texture of the neutrino mass matrix. 
We will refer to them as the $S_3$ conditions.

\section{Minimal seesaw model}

In the minimal seesaw model (as reviewed, for example, in Ref.\,\cite{GXZ})
the neutrino mass matrix is written as
\begin{equation}\label{Mseesaw}
M_{\nu}\,=\,m_D\,M_R^{-1}m_D^T
\end{equation}
where $m_D$ is a $3\times\,2$ matrix
\begin{equation}\label{M32}
m_D\,=\,\left(\begin{array}{cc}
                     a_1 & a_2  \\
                     b_1 & c_1  \\
                     b_2 & c_2  \end{array} \right)\,,
\end{equation}
and $M_R$ is a $2\times\,2$ matrix,
\begin{equation}\label{MR}
M_R\,=\,\left(\begin{array}{cc}
                         M_{22} & M_{23}  \\
                         M_{23} & M_{33}  \end{array} \right)\,.
\end{equation}
The rank of $M_{\nu}$ is two and therefore one of the eigenvalues 
of (\ref{Mseesaw}) must be zero.

Evaluating (\ref{Mseesaw}) for the parameters in (\ref{M}) gives
\begin{eqnarray}
A\,&=&\,\frac{a_2^2M_{22}-2a_1a_2M_{23}+a_1^2M_{33}}{\mathcal{D}}  \label{Ass}  \\
\frac{1}{2}(B_1+B_2)\,&=&\,\frac{a_2(c_1+c_2)M_{22}-[a_2(b_1+b_2)+a_1(c_1+c_2)]M_{23}+a_1(b_1+b_2)M_{33}}{2\mathcal{D}} \label{B1ss}  \\
\frac{1}{2}(C_1+C_2)\,&=&\,\frac{(c_1^2+c_2^2)M_{22}-2(b_1c_1+b_2c_2)M_{23}+(b_1^2+b_2^2)M_{33}}{2\mathcal{D}}  \label{Css}  \\
D\,&=&\,\frac{c_1c_2M_{22}-(b_2c_1+b_1c_2)M_{23}+b_1b_2M_{33}}{\mathcal{D}}  \label{Dss}  \\
\frac{1}{2}(B_1-B_2)\,&=&\,\frac{a_2(c_1-c_2)M_{22}-[a_2(b_1-b_2)+a_1(c_1-c_2)]M_{23}+a_1(b_1-b_2)M_{33}}{2\mathcal{D}} \label{dBss}  \\
\frac{1}{2}(C_1-C_2)\,&=&\,\frac{(c_1^2-c_2^2)M_{22}-2(b_1c_1-b_2c_2)M_{23}+(b_1^2-b_2^2)M_{33}}{2\mathcal{D}}  \label{dCss}
\end{eqnarray}
where $\mathcal{D}\,=\,M_{22}M_{33}-M_{23}^2$ is the determinant of (\ref{MR}).
Now using the $S_3$ conditions (\ref{S1}) and (\ref{S2}) we get, after a lot of simplification,
\begin{eqnarray}
0\,&=&\,(2a_2-c_1-c_2)\left[(a_2+c_1+c_2)\frac{M_{22}}{\mathcal{D}}-(a_1+b_1+b_2)\frac{M_{23}}{\mathcal{D}}\right] \nonumber  \\
   &-&\,(2a_1-b_1-b_2)\left[(a_2+c_1+c_2)\frac{M_{23}}{\mathcal{D}}-(a_1+b_1+b_2)\frac{M_{33}}{\mathcal{D}}\right]\,, \label{see1} \\
0\,&=&\,(c_1-c_2)\left[(a_2+c_1+c_2)\frac{M_{22}}{\mathcal{D}}-(a_1+b_1+b_2)\frac{M_{23}}{\mathcal{D}}\right] \nonumber  \\
   &-&\,(b_1-b_2)\left[(a_2+c_1+c_2)\frac{M_{23}}{\mathcal{D}}-(a_1+b_1+b_2)\frac{M_{33}}{\mathcal{D}}\right]\,. \label{see2}
\end{eqnarray}
Since we assume that $b_1\ne\,b_2,\,\,c_1\ne\,c_2$ (to avoid $B_1=B_2,\,\,\, C_1=C_2$), Eqs. (\ref{see1}) and (\ref{see2}) could be
solved by requiring
\begin{eqnarray}
a_1+b_1+b_2\,&=&\,0\,,\label{abc1}  \\
a_2+c_1+c_2\,&=&\,0\,.\label{abc2}
\end{eqnarray}
But, when we put these relations back into (\ref{Ass}) - (\ref{dCss}), we get additional unwanted constraints
\begin{eqnarray}
B_1+B_2\,&=&\,-\,A\,, \label{newABB}  \\
C_1+C_2+2\,D\,&=&\,A\,. \label{newCCDA}
\end{eqnarray}
Thus the solution of (\ref{see1}) and (\ref{see2}) must involve conditions on the parameters of $M_R$ as well as those of $m_D$.

Another way to understand the $S_3$ conditions and the restrictions (\ref{abc1}), (\ref{abc2})
is to consider a $Z_2$ symmetry \cite{SHY,DGH}
\begin{equation}\label{Z1}
G_1^T\,M_{\nu}G_1\,=\,M_{\nu}
\end{equation}
where
\begin{equation}\label{Z2}
G_1\,=\,\frac{1}{2+k^2}\left(\begin{array}{ccc}
                               2-k^2 & 2k & 2k  \\
                                2k & k^2 & -2  \\
                                2k & -2 & k^2  \end{array}\right).
\end{equation}
Eq.\,(\ref{Z1}) gives two conditions
\begin{eqnarray}
\frac{B_1+B_2}{C_1+C_2+2D-2A}\,&=&\,\frac{k}{k^2-2}  \label{Z3}  \\
\frac{B_1-B_2}{C_1-C_2}\,&=&\,\frac{1}{k}  \label{Z4}
\end{eqnarray}
which are the $S_3$ conditions if $k=-1$.

In the following sections we turn to finding restrictions on the mixing angles.  In those sections the only use 
we will make of the minimal seesaw model 
is as motivation for setting one mass eigenvalue equal to zero.


\section{Eigenvalues and eigenstates}

For a zero mass eigenvalue we have, 
\begin{equation}\label{m00}
\left(\begin{array}{ccc}
           A & B_1 & B_2  \\
           B_1 & C_1 & D  \\
           B_2 & D & C_2 \end{array}\right)\,\left(\begin{array}{c}
                                                       \alpha  \\
                                                       \beta  \\
                                                       \gamma  \end{array} \right)\,=\,
                                       \lambda\left(\begin{array}{c}
                                                       \alpha  \\
                                                       \beta   \\
                                                       \gamma  \end{array} \right)\,=\,0\,.
\end{equation}
If we assume $\alpha\,\ne\,0$ (we will check this below) then we get three equations
\begin{eqnarray}
A\,&=&\,-\rho\,B_1-\sigma\,B_2  \label{ABB}  \\
B_1\,&=&\,-\rho\,C_1-\sigma\,D  \label{BCD}  \\
B_2\,&=&\,-\rho\,D-\sigma\,C_2  \label{BDC}
\end{eqnarray}
where $\rho\equiv\beta/\alpha,\,\sigma\equiv\gamma/\alpha$.    So $B_1,B_2$ and $A$ are given by (\ref{BCD}), (\ref{BDC}), and
\begin{equation}\label{AA}
A\,=\,\rho^2C_1+\sigma^2C_2+2\rho\sigma\,D
\end{equation}
Now let's use these in the $S_3$ relations.  Eq.\,(\ref{S2}) and Eq.\,(\ref{S1}) give
\begin{eqnarray}
(\sigma-1)C_2-(\rho-1)C_1+(\rho-\sigma)D=0  \label{C2C1D}  \\
(2\rho^2-\rho-1)C_1+(2\sigma^2-\sigma-1)C_2+(4\rho\sigma-\rho-\sigma-2)D=0  \label{C1C2D}
\end{eqnarray}
Eqs.\,(\ref{C2C1D}) and (\ref{C1C2D}) can be reduced to
\begin{eqnarray}
(\rho+\sigma+1)[(1-\sigma)C_2+(1-\rho)D]\,&=&0\,,\label{newC2D}  \\
(\rho+\sigma+1)[(1-\rho)C_1+(1-\sigma)D]\,&=&0\,, \label{newC1D}
\end{eqnarray}
If we choose the solutions
\begin{eqnarray}
C_1\,&=&\,-\frac{1-\sigma}{1-\rho}D\,, \label{solC1D} \\
C_2\,&=&\,-\frac{1-\rho}{1-\sigma}D\,, \label{solC2D}
\end{eqnarray}
then $D^2=C_1C_2$.  The mass eigenvalues that we expect to be nonzero are given by
\begin{equation}\label{newmpmm}
m_{\pm}\,=\,\frac{1}{2}\left[A+C_1+C_2\pm\sqrt{(A+C_1+C_2)^2+4(\rho^2+\sigma^2+1)(D^2-C_1C_2)}\right]
\end{equation}
where we have used (\ref{BCD}) and (\ref{BDC}).  Thus this solution
makes a second mass eigenvalue zero.  We need nonzero two masses in order to have two oscillation lengths.
We might tolerate two zero masses in the case of normal hierarchy, $m_3\gg\,m_2\approx\,m_1$.
We will ignore that special case except for a brief comment at the end of Sec. 5.
Thus the only way to avoid two zero masses is to require $\rho+\sigma+1\,=\,0$.  

If we take $C_1$ and $C_2$ as the independent variables the nonzero eigenvalues are,
from (\ref{newmpmm}) using (\ref{AA}),
\begin{equation}\label{pmmp}
m_{\pm}\,=\,\frac{(2+2\sigma-\sigma^2)C_1+(1+4\sigma+\sigma^2)C_2}{2(2\sigma+1)}
\pm\frac{3}{2}\left|\frac{(2+2\sigma+\sigma^2)C_1-(1+\sigma^2)C_2}{2\sigma+1}\right|
\end{equation}
The total set of eigenvalues and eigenfunctions can be reduced to
\begin{eqnarray}
m_0\,&=&0\,,  \label{m0}  \\
|\nu_0>\,&=&\,\frac{1}{\sqrt{2}\sqrt{1+Re(\sigma)+|\sigma|^2}}[|\nu_e>-(1+\sigma)|\nu_{\mu}>+\sigma|\nu_{\tau}>]  \label{n0}  \\
m_{a}\,&=&\,\frac{1}{2\sigma+1}[(\sigma+2)^2\,C_1-(\sigma-1)^2\,C_2]  \label{mp}  \\
|\nu_{a}>\,&=&\,\frac{1}{\sqrt{3}}[|\nu_e>+|\nu_{\mu}>+|\nu_{\tau}>]  \label{np}  \\
m_{b}\,&=&\,\frac{2(\sigma^2+\sigma+1)}{2\sigma+1}[C_2-C_1]  \label{mm}  \\
|\nu_{b}>\,&=&\,\frac{1}{\sqrt{6}\sqrt{1+Re(\sigma)+|\sigma|^2}}[-(1+2\sigma)|\nu_e>- (1-\sigma)|\nu_{\mu}>+(2+\sigma)|\nu_{\tau}>]
                                                                                                             \label{nm}
\end{eqnarray}
where $a,b$ are $+,-$ or $-,+$ depending on the signs of the factors in the absolute value of (\ref{pmmp}).

The only remaining case is to go back to (\ref{m00}) and set $\alpha\,=\,0$.  Eqs.\,(\ref{ABB}), (\ref{BCD}), and (\ref{BDC})
are then
\begin{eqnarray}
B_2\,&=&\,-\lambda\,B_1  \label{alp1} \\
D\,&=&\,-\lambda\,C_1  \label{alp2} \\
C_2\,&=&\,-\lambda\,D\,=\,\lambda^2\,C_1  \label{alp3}  
\end{eqnarray}
where $\lambda\,\equiv\,\beta/\gamma$.  The $S_3$ conditions are 
\begin{eqnarray}
B_1\,&=&\,(\lambda-1)C_1  \label{alp4} \\
A\,&=&\,(\lambda-1)^2\,C_1\,. \label{alp5}
\end{eqnarray}
Since one eigenvalue is zero the remaining eigenvalues are given by
\begin{equation}\label{lammas}
\lambda_{\pm}\,=\,\frac{1}{2}\left[A+C_1+C_2\pm\sqrt{(A+C_1+C_2)^2+4(D^2+B_1^2+B_2^2-C_1C_2-AC_1-AC_2)}\right]
\end{equation}
and (\ref{alp1}) - (\ref{alp5}) give $D^2+B_1^2+B_2^2-C_1C_2-AC_1-AC_2\,=\,0$.
Thus $\alpha\,=\,0$ would require a second mass eigenvalue to be zero.

So the only solution with broken $\mu-\tau$ symmetry and two nonzero masses is given by (\ref{m0}) - (\ref{nm}).
Since we know $\mu-\tau$ symmetry is approximately true the parameter $\sigma$ will need to be large
for inverted hierarchy or approximately $-\frac{1}{2}$ for normal hierarchy.

\section{Restrictions on the mixing angles}

The conditions on the elements of the mass matrix $A,\ldots\,,D$
will allow us to put conditions on the mixing angles $(\theta_s,\theta_a,\theta_x)\,\equiv\,(\theta_{12},\theta_{23},\theta_{13})$.
The neutrino mixing matrix \cite{19} which diagonalizes $M_{\nu}$ via $V^TM_{\nu}V\,=\,M_{\nu}^{{\rm diag}}$
can be decomposed as $V=U''UU'$ \cite{SHY} where $U$ is a CKM type matrix
\begin{equation}\label{U}
U\,=\,\left(\begin{array}{ccc}
            c_sc_x & -s_sc_x & -s_xp  \\
            s_sc_a-c_ss_as_xp^{*} & c_sc_a+s_ss_as_xp^{*} & -s_ac_x   \\
            s_ss_a+c_sc_as_xp^{*} & c_ss_a-s_sc_as_xp^{*} & c_ac_x  \end{array} \right)
\end{equation}
with $(s_{\alpha},c_{\alpha})\,\equiv\,(\sin\theta_{\alpha},\cos\theta_{\alpha})$ for $\alpha\,=\,s,a,x$
and $p=e^{i\delta_D}$ where $\delta_D$ is the Dirac phase, 
$U''$ is the rephasing matrix, ${\rm diag}(e^{i\alpha_1},e^{i\alpha_2},e^{i\alpha_3})$\cite{BM},
and
$U'\,=\,{\rm diag}(e^{-i\phi_1/2},e^{-i\phi_2/2},e^{-i\phi_3/2})$ 
where $\phi_1$, $\phi_2$, and $\phi_3$ are Majorana phases.
The neutrino mass matrix is then
\begin{equation}\label{VMV}
M_{\nu}\,=\,V^{*}M_{\nu}^{{\rm diag}}V^{\dagger}
\end{equation}
with elements given by
\begin{eqnarray}
A\,&=&\,\left[c_x^2c_s^2m_1'+c_x^2s_s^2m_2'+p^{*2}s_x^2m_3'\right]e^{-2i\alpha_1}   \label{VA}  \\
B_1\,&=&\,\left[c_x[s_sc_sc_a-ps_xs_ac_s^2]m_1'-c_x[s_sc_sc_a+ps_xs_as_s^2]m_2'+p^{*}c_xs_xs_am_3'\right]
e^{-i(\alpha_1+\alpha_2)}  \label{VB1}  \\
B_2\,&=&\,\left[c_x[s_sc_ss_a+ps_xc_ac_s^2]m_1'-c_x[s_sc_ss_a-ps_xc_as_s^2]m_2'-p^{*}s_xc_xc_am_3'\right]
e^{-i(\alpha_1+\alpha_3)}  \label{VB2}  \\
C_1\,&=&\,\left[(s_sc_a-ps_xc_ss_a)^2m_1'+(c_sc_a+ps_xs_ss_a)^2m_2'+c_x^2s_a^2m_3'\right]
e^{-2i\alpha_2}  \label{VC1}  \\
C_2\,&=&\,\left[(s_ss_a+ps_xc_sc_a)^2m_1'+(c_ss_a-ps_xs_sc_a)^2m_2'+c_x^2c_a^2m_3'\right]
e^{-2i\alpha_3}  \label{VC2}  \\
D\,&=&\,\big[(s_ss_a+ps_xc_sc_a)(s_sc_a-ps_xc_ss_a)m_1'   \nonumber  \\
   &+&\,(c_ss_a-ps_xs_sc_a)(c_sc_a+ps_xs_ss_a)m_2'-c_x^2s_ac_am_3'\big]e^{-i(\alpha_2+\alpha_3)}   \label{VD}
\end{eqnarray}
where $m_1'=m_0\sqrt{1+r}e^{i\phi_1},\,m_2'=m_0e^{i\phi_2},\,m_3'=0$ for inverted hierarchy
or $m_1'=0,\, m_2'=m_0\sqrt{r}e^{i\phi_2},\,m_3'=m_0\sqrt{1+r}e^{i\phi_3}$ for normal hierarchy.
This $m_0$ is a universal mass, not the same as the eigenvalue
of the previous section, and $r$ is the ratio of the mass splittings, $r\equiv\,\Delta_s/\Delta_a$.
It is easy to see that $\mu-\tau$ symmetry requires $s_a=c_a$ and $s_x=0$.  Tri-bimaximal symmetry requires, in addition,
$\tan\theta_s\,=\,\sqrt{2}\,\,\,{\rm or}\,-1/\sqrt{2}$.

We have four relations for $\sigma$ from (\ref{ABB}), (\ref{BCD}), (\ref{BDC}), and (\ref{C2C1D}), all with
$\rho$ replaced by $-\sigma-1$.  The first three are
\begin{eqnarray}
\frac{1}{\sigma}\,&=&\,\frac{B_1-B_2}{A-B_1}\,,  \label{sig1}  \\
\frac{1}{\sigma}\,&=&\,\frac{C_1-D}{B_1-C_1}\,,  \label{sig2}  \\
\frac{1}{\sigma}\,&=&\,\frac{D-C_2}{B_2-D}\,.   \label{sig3}
\end{eqnarray}
Now we assume inverted hierarchy and substitute (\ref{VA}) - (\ref{VD}) on the RHSs to find relations on the mixing angles. 
Only two of these are independent because we have set $m_3'$ equal to zero so we get just one relation
among the angles,
\begin{equation}\label{r1}
s_x\,e^{i\delta_D}\,=\,c_x\left[c_a\,e^{i(\alpha_3-\alpha_1)}-s_a\,e^{i(\alpha_2-\alpha_1)}\right]\,,
\end{equation}
and the solution for $\sigma$,
\begin{equation}\label{r2}
\frac{1}{\sigma}\,=\,\frac{s_a\,e^{i(\alpha_2-\alpha_3)}-c_a}{c_a}\,.
\end{equation}
One of the mass eigenvalues $m_a,\,m_b$, given by (\ref{mp}) and (\ref{mm}), must equal $m'_1$ and the other $m'_2$. 
If we evaluate $m_a$ and $m_b$ using (\ref{r2}) and $C_1,\,C_2$ given by (\ref{VC1}), (\ref{VC2}) we find this requires
$\alpha_2=\alpha_3$.  Thus $\sigma$ is real and (\ref{r1}) requires
\begin{equation}\label{three}
\delta_D+\alpha_1-\alpha_2\,=\,0\,\,{\rm or}\,\,\pi\,.
\end{equation}
It is immaterial which sign we take from (\ref{three});
in what follows we use
\begin{eqnarray}
s_x\,&=&\,c_x(c_a-s_a)  \label{X1}  \\
\frac{1}{\sigma}\,&=&\,\frac{s_a-c_a}{c_a}  \label{X2}
\end{eqnarray}

We still have the condition from $S_3$, Eq.\,(\ref{C2C1D}), which now depends only on $\delta_D$,
\begin{equation}\label{sig4}
\frac{1}{\sigma}\,=\,\frac{2D-C_1-C_2}{2C_1-C_2-D}\,.
\end{equation}
Using (\ref{X2}) for the LHS and (\ref{r1}) for $e^{i\delta_D}$, this gives a quadratic equation for $\tan\theta_s$
\begin{eqnarray}
&&\tan\theta_s\,=\,\frac{m_1e^{-i\alpha}-m_2\,e^{i\alpha}}{(s_a+c_a)(m_1^2+m_2^2-2m_1m_2\cos2\alpha)} \nonumber \\
      &&\left[c_x(1-4c_as_a)(m_1-m_2)\pm\sqrt{c_x^2(1-4c_as_a)^2(m_1-m_2)^2+4(1+2\,c_a\,s_a)(m_1^2+m_2^2-2m_1m_2\cos\,2\alpha)}\right]
\label{TAN}
\end{eqnarray}
where $\alpha\equiv\alpha_1-\alpha_2$.
This gives real values only if $\delta_D$ is zero or $\pi$.
One solution of the quadratic equation is
\begin{equation}\label{r3}
\tan\theta_s\,=\,-\frac{2c_x}{c_a+s_a}[1-s_ac_a]
\end{equation}
or, when we use (\ref{X1}),
\begin{equation}\label{r3p}
\tan\theta_s\,=\,-\frac{1}{c_x(s_a+c_a)}\,.
\end{equation}
The other solution is 
\begin{equation}\label{r3a}
\tan\theta_s\,=\,\frac{c_x}{c_a+s_a}[1+2c_as_a]\,=\,c_x(c_a+s_a)\,.
\end{equation}
Since $\tan(\frac{\pi}{2}-\theta_s)\,=\,1/\tan\theta_s$, and since the oscillation experiments measure $\sin^2\,2\theta$,
these two solutions are effectively equivalent.

Another way of expressing the results is to write all of the mixing angles in terms of the one parameter $\sigma$.
From (\ref{X2}) and then (\ref{X1}) we find
\begin{eqnarray}
\tan\theta_a\,&=&\,\frac{\sigma+1}{\sigma}\,,  \label{r4}  \\
\tan\theta_x\,&=&\,\frac{-1}{\sqrt{1+2\sigma+2\sigma^2}}\,,  \label{r5}  
\end{eqnarray}
and, from (\ref{r3}),
\begin{eqnarray}
\tan\theta_s\,&=&\,-\frac{\sqrt{2}\sqrt{1+\sigma+\sigma^2}}{1+2\sigma}  \label{r6}
\end{eqnarray}
or, from (\ref{r3a}),
\begin{eqnarray}
\tan\theta_s\,&=&\,\frac{1+2\sigma}{\sqrt{2}\sqrt{1+\sigma+\sigma^2}}\,.  \label{r6a}
\end{eqnarray}
These two solutions have $m_{b}=m_1'=m_0\sqrt{1+r},\,\,m_{a}=m_2'=m_0$ for (\ref{r6}) or
$m_{b}=m_2'=m_0,\,\,m_{a}=m_1'=m_0\sqrt{1+r}$ for (\ref{r6a}).
(Here we are only concerned with the magnitude of the masses so we have neglected Majorana phases.)
Since $|\sigma|$ is large, Eq.\,(\ref{nm}) shows that $|\nu_b>$ has a larger fraction of $|\nu_e>$ than does $|\nu_a>$
so $m_b$ should be smaller than $m_a$.  If $r\,>\,0$ then Eq.\,(\ref{r3a}) gives $m_a=m_0\sqrt{1+r}\,>\,m_b=m_0$
but Eq.\,(\ref{r3p}) would require $r\,<\,0$.
However, as mentioned above, these two solutions for $\tan\theta_s$ are effectively indistinguishable so we won't worry about this point further.

So far we have considered only inverted hierarchy but
Eq.\,(\ref{n0}) supports normal hierarchy with $\sigma\approx-\frac{1}{2}$.
If we set $m_1=0$ and compare the right hand sides of (\ref{sig1}), (\ref{sig2}), and (\ref{sig3}) we get the condition
\begin{equation}\label{rr1}
c_s(c_x-s_x\,(s_a-c_a))+s_s(s_a+c_a)\,=\,0\,,
\end{equation}
where to make the algebra simplier we immediately neglect all the phases.
This expression for $\tan\theta_s$,
\begin{eqnarray} 
\tan\theta_s\,=\,-\frac{c_x-s_x(s_a-c_a)}{s_a+c_a}\,,  \label{rr3}
\end{eqnarray}
superfically looks different than those of the inverted case.
However, if we now set (\ref{sig4}) equal to (\ref{sig1}) or (\ref{sig2}) or (\ref{sig3}) we find Eq.\,(\ref{X1}) which,
when combined with (\ref{rr3}), reproduces our first solution for $\tan\theta_s$ in the inverted hierarchy case, Eq.\,(\ref{r3p}).

Using these conditions the solution for $\sigma$ is now
\begin{equation}\label{rr5}
\sigma\,=\,\frac{c_a-2s_a}{s_a+c_a}
\end{equation}
or
\begin{equation}\label{nta}
\tan\theta_a\,=\,\frac{1-\sigma}{2+\sigma}
\end{equation}
When we solve for $m_a,\,m_b$ of Eqs.\,(\ref{mp}), (\ref{mm}) we get
\begin{eqnarray}
m_{b}\,&=&\,m_3\,=\,m_0\sqrt{1+r} \label{nhmm}  \\
m_{a}\,&=&\,m_2\,=\,m_0\sqrt{r}  \label{nhpm}
\end{eqnarray}
If we set $m_2=0$ rather than $m_1$, we get the second expression for $\tan\theta_s$ of inverted hierarchy, (\ref{r3a}),
and $m_a=m_1$.
For this normal hierarchy case $\sigma$ is approximately $-\frac{1}{2}$ and the eigenfunction with zero mass, Eq.\,(\ref{n0}), 
has the largest fraction of $|\nu_e>$ as it should.

We have insisted that only one mass eigenvalue be zero.  If two masses are zero then, from (\ref{VA}) - (\ref{VD}) with only $m'_3$ nonzero,
the $S_3$ conditions are both satisfied by (\ref{r1}) above.

\section{Summary}
We assume that the neutrino mass matrix is given by the three dimensional representation of the group $S_3$
and that one (but only one\,!) of the mass eigenvalues is zero, as required, for example, by the minimal seesaw model.
The experimentally testable results for either inverted or normal hierarchy a real
Dirac phase factor
(there could be nonzero Majorana phases but they don't affect our results),
\begin{eqnarray}
\delta_D&=&0\,\,,\pi  \label{r7}  
\end{eqnarray}
and two conditions on the mixing angles,
\begin{eqnarray}
\tan\theta_x\,&=&\,c_a-s_a\,,   \label{r9}
\end{eqnarray}
and
\begin{eqnarray}
\tan\theta_s\,&=&\,-\frac{1}{c_x(s_a+c_a)}  \label{r10a}
\end{eqnarray}
or
\begin{equation}\label{r10b}
\tan\theta_s\,=\,c_x(s_a+c_a)\,.
\end{equation}

For inverted hierarchy we can convert (\ref{r4}) - (\ref{r6a}) to 
\begin{eqnarray}
\sin^2\theta_a\,&=&\,\frac{(1+\sigma)^2}{1+2\sigma+2\sigma^2}\,, \label{sina}  \\
\sin^2\theta_x\,&=&\,\frac{1}{2(1+\sigma+\sigma^2)}\,, \label{sinb}  
\end{eqnarray}
and
\begin{equation}\label{sin1}
\sin^2\theta_s\,=\,\frac{2(1+\sigma+\sigma^2)}{3(1+2\sigma+2\sigma^2)}
\end{equation}
or
\begin{equation}\label{sin2}
\sin^2\theta_s\,=\,\frac{(1+2\sigma)^2}{3(1+2\sigma+2\sigma^2)}\,.
\end{equation}
to more easily compare with the data.  The result of fitting these expressions to the
experimental values \cite{FLMPR,Fetal} is shown in the following table.   
As mentioned above
the oscillation expressions depend on $\sin^22\theta$ and thus the experiments can't distinguish between $\theta_s$ greater
or less than $\pi/4$.  So the fit using (\ref{sin1}) assumes the experimental value is less than $\pi/4$,
while that using (\ref{sin2}) assumes it is greater than $\pi/4$, since these are the values indicated by the formula.
(Approximate $\mu-\tau$ symmetry gives $s_a\sim\,c_a\sim\,\frac{1}{\sqrt{2}}$ so (\ref{r10a}) gives $\tan\theta_s\sim\frac{1}{\sqrt{2}}$
while (\ref{r10b}) gives $\tan\theta_s\sim\sqrt{2}$; thus a fit of (\ref{sin2}) with $\theta_s\,<\,\pi/4$ is untenable.)
The two sets of expressions give equivalent fits to the data as they must.


\begin{center}
\begin{table}[h]
\begin{tabular}{|c|c|c|c|c|}\hline
Angles & Best Fit & Exp Range & Fit 1 & Fit 2  \\ \hline  
$\sin^2\theta_a$ & $0.466$ & $0.408 - 0.539$ & $0.425$ & $0.425$  \\ \hline
$\sin^2\theta_x$ & $0.016$ & $0.006 - 0.026$ & $0.011$ & $0.011$   \\ \hline  
$\sin^2\theta_s$ & $0.312$ & $0.294 - 0.331$ & $0.337$ & $--$  \\ \hline
$\sin^2\theta_s$ & $0.688$ & $0.669 - 0.706$ & $--$ & $0.663$  \\ \hline 
\end{tabular}
\caption {The second column gives the experimental best fit, the third column gives the $1-\sigma$ experimental range, 
the fourth column gives the central values using (\ref{sin1}), the fifth column gives the central values using (\ref{sin2}). 
The minimum $\chi^2$ is $2.15$.
The value of $\sigma$ which gives the minimum is $-7.17$ if we use (\ref{X2}) or $-0.388$ if we use (\ref{rr5}).
Our fits give values for the angles of $|\theta_a|\,=\,40.7^{\circ},\,\,|\theta_x|\,=\,6.02^{\circ},\,\,$ and 
$|\theta_s|\,=\,35.5^{\circ}$ or $54.5^{\circ}$.}
\end{table}
\end{center}   

Similarly for a normal hierarchy of masses we could find expressions for $\sin^2\theta_a$, $\sin^2\theta_x$,
and $\sin^2\theta_s$ in terms of the $\sigma$ for normal hierarchy given by (\ref{rr5}).  
But if we call that $\sigma_N$, and the $\sigma$ given by (\ref{X2}) $\sigma_I$, then (\ref{r4}) and (\ref{nta}) give
\begin{equation}\label{sigmas}
\sigma_N\,=\,-\frac{2+\sigma_I}{1+2\sigma_I}
\end{equation}
and if we replaced $\sigma_N$ by $\sigma_I$ in normal hierarchy expressions for $\sin^2\theta_a, \sin^2\theta_x, \sin^2\theta_s$
we would reproduce (\ref{sina}) - (\ref{sin2}).
The normal hierarchy expressions would be just 
a reparameterization of (\ref{sina}) - (\ref{sin2}) above and thus give an identical fit.

Recently MINOS\cite{MINOS} has presented a measurement of $\sin^2(2\theta_a)\,\sin^2(2\theta_x)$.
We can easily fit this assuming the values above for $\sin^2\theta_a$ and $\sin^2\theta_s$
but replacing $\sin^2\theta_x$ by this combination.  This is shown in Table II.
The fact that the fit values of $\sin^2\theta_x$ are smaller than the value in Table I despite the experimental number being
bigger is because of the larger error in that number.  Still as $\chi^2$ increases by one from its minimum, $\sin\theta_x$ varies
from $0.$ to only $0.024$, which implies that this model prefers small $\theta_x$.
\begin{center}
\begin{table}[h]
\begin{tabular}{|c|c|c|c|c|}\hline
Angles & Best Fit & Exp Range & IH Fit & NH Fit  \\ \hline
$\sin^2\theta_a$ & $0.466$ & $0.408 - 0.539$ & $0.449$ & $0.443$  \\ \hline
$\sin^2(2\theta_a)\sin^2(2\theta_x)$ & $0.18$ & $0.06 - 0.32$ & $0.021$ & $--$  \\ \hline
$\sin^2(2\theta_a)\sin^2(2\theta_x)$ & $0.11$ & $0.04 - 0.21$ & $--$ & $0.025$  \\ \hline
$\sin^2\theta_s$ & $0.312$ & $0.294 - 0.331$ & $0.335$ & $0.335$ \\ \hline
Minimum $\chi^2$ & & & $2.90$ & $2.37$  \\ \hline
\end{tabular}
\caption {The fourth and fifth columns give the results of fitting the MINOS values given in row two
or row three. Again the experimental range is $1-\sigma$. The MINOS numbers depend on whether they assume inverted or normal hierarchy.
The fitted numbers correspond to a $\sin^2\theta_x$ of $0.0052$ for IH or $0.0064$ for NH.
The values of the angles are therefore $|\theta_a|\,=\,42.1^{\circ}$ or $41.7^{\circ}$, 
$|\theta_x|\,=\,4.14^{\circ}$ or $4.59^{\circ}$, and $|\theta_s|\,=\,35.4^{\circ}$.}
\end{table}
\end{center}

\section*{Acknowledgments}
SFG was supported by the Chinese Scholarship Council. DAD and SFG were supported in part by the U. S. Department of Energy under grant No. DE-FG03-93ER40757. WWR was supported in part by the National Science Foundation under Grant PHY-0555544. We thank Sacha Kopp for discussions of the MINOS results. DAD is a member of the Center for Particles and Fields and the Texas Cosmology Center.
\appendix
\section{Charged Lepton and Neutrino Sectors}

In recent years it has become common to attempt a unified treatment of the charged lepton sector and the neutrino sector. In this paper, we discuss only the neutrino sector, as do many of our references.  This appendix shows that the sectors can be discussed separately.

The essential point of flavor mixing is that the physical mixing matrix, being misaligned between two representations of the up-type and down-type fermions,  is independent of formalism or representation. For the lepton sector, the physical mixing matrix is the so-called PMNS matrix \cite{19},
\begin{equation}
  V_{PMNS}=U^\dagger_e U_\nu\,,
\end{equation}

\noindent
where the two flavor mixing matrices are determined by
\begin{equation}
  U^\dagger_e M_e M^\dagger_e U_e=D_e D^\dagger_e\,,
\qquad U^T_\nu M_\nu U_\nu = D_\nu\,.
  \label{eq:diag}
\end{equation}

\noindent
As in the body of the paper, we take the neutrinos to be Majorana particles. The diagonal mass matrices are denoted as $D_e$ and $D_\nu$ for charged leptons and neutrinos. Now, we can make an arbitrary rotation on all the lepton fields, including left-handed charged leptons and neutrinos as well as the right-handed charged leptons. Since the left-handed charged leptons and neutrinos reside in common $SU(2)_L$ doublets, they share a common rotation,
\begin{equation}
  \begin{pmatrix}
    \nu_i \\
    \ell_i
  \end{pmatrix}_L
\rightarrow
  (T_L)_{ij}
  \begin{pmatrix}
    \nu_j \\
    \ell_j
  \end{pmatrix}_L,
\qquad
  (\ell_i)_R
\rightarrow
  (T_R)_{ij} 
  (\ell_j)_R\,.
\end{equation}

\noindent
Then the charged lepton and neutrino mass matrices become
\begin{equation}
  M_e\rightarrow\widetilde M_e=T_L M_e T^\dagger_R\,,
\qquad M_\nu\rightarrow\widetilde M_\nu=T^*_L M_\nu T^\dagger_L\,,
\end{equation}

\noindent
and we denote the modified mixing matrices as $\widetilde U_e$ and $\widetilde U_\nu$ respectively. Eq.(\ref{eq:diag}) becomes
\begin{equation}
  \widetilde U^\dagger_e \widetilde M_e \widetilde M^\dagger_e \widetilde U_e
= D_e D^\dagger_e\,, \qquad \widetilde U^T_\nu \widetilde M_\nu \widetilde U_\nu=D_\nu\,,
\end{equation}

\noindent
where
\begin{equation}
  \widetilde U_e =T_L U_e\,,\qquad\widetilde U_\nu=T_L U_\nu\,.
\end{equation}
The important point is that the physical mixing matrix is not affected,
\begin{equation}
  \widetilde V_{PMNS}=\widetilde U^\dagger_e \widetilde U_\nu=
  U^\dagger_e T^\dagger_L T_L U_\nu=U^\dagger_e U_\nu=V_{PMNS}\,.
\end{equation}

\noindent
This is expected because, if it were not true, the physical mixing matrix would depend on the formalism or representation. This property of formalism/representation independence allows us to rotate the charged leptons to a mass diagonal basis, since what we want to discuss is just the physical mixing matrix. It doesn't matter in which basis the discussion is made. The question is how to realize this.

We should also note that, after gauge symmetry breaking where the fermions acquire mass, the up-type and down-type fermions' mass matrices should be constrained by different representations of some symmetry if there is any. Otherwise, the two mass matrices would be constrained to be of the same form and this would lead to trivial physical mixing. The full group is at least a product.

We can imagine that the $S_3$ discussed in current work is kind of residual property of some symmetry. Before symmetry breaking, there would be a larger group governing both the charged lepton and neutrino sectors, especially the left-handed ones since they share a common left-handed doublet. But experimentally the symmetry is broken. There can be a residual symmetry for the left-handed charged leptons, for example ${\cal Z}_3$,
\begin{equation}
  {\cal Z}_3 = \{ I, F, F^2 \}\quad \mbox{with} \quad F=\begin{pmatrix}
                                                      1 \\
                                                      & \omega \\
                                                      & & \omega^2
                                                 \end{pmatrix}\,,
\end{equation}

\noindent
where $\omega \equiv e^{2 i \pi / 3}$. If we use $\mathcal F$ to denote the group elements of ${\cal Z}_3$, then the charged lepton's mass matrix has to satisfy
\begin{equation}
  \mathcal F^\dagger M_e M^\dagger_e \mathcal F=D_e D^\dagger_e\,.
\end{equation} 

\noindent
It can be verified that under this constraint $M_e M^\dagger_e$ has to be diagonal,
\begin{equation}
  M_e M^\dagger_e=D_e D^\dagger_e\,,
\end{equation}

\noindent
and $U_e = I$. In other words, the physical mixing comes solely from the neutrino sector. Actually, ${\cal Z}_3$ is a subgroup of $S_3$ so we can  apply $S_3$ in the charged lepton sector. But, as argued above, the representations of the charged lepton and neutrino sectors should be different. In other words, the residual ${\cal Z}_3$ of the charged lepton sector cannot be simply embodied in the residual $S_3$ of the neutrino sector. The unified group should be at least a product group \cite{Lam}
\begin{equation}
{\cal G}={\cal Z}_3\otimes S_3\,.
\end{equation}
If we want to apply $S_3$ in the charged lepton sector too, this can be
achieved by embedding ${\cal Z}_3$ in another $S_3$ whose representation is
different from that of neutrino sector. Then the product group would
be $S_3 \otimes S_3$.


\begin{thebibliography}{99}
\bibitem{intro}S. Pakvasa and H. Sugawara, Phys. Lett. B{\bf 73}, 61 (1978); {\bf 82}, 105 (1979);
E. Durman and H.S.Tsao, Phys. Rev. D{\bf 20}, 1207 (1979);
Y. Yamanaka, H. Sugawara, and S. Pakvasa, Phys. Rev. D{\bf 25}, 1895 (1982);
K. Kang, J. E. Kim, and P. Ko, Z. Phys. C{\bf 72}, 671 (1996), hep-ph/9503346;
K. Kang, S. K. Kang, J. E. Kim, and P. Ko, Phys. Lett. A{\bf 12}, 1175 (1996), hep-ph/9611396; 
M. Fukugita, M. Tanimoto, and T. Yanagida, Phys. Rev. D{\bf 57}, 44299 (1998), hep-ph/9709388;
H. Fritzsch and Z-z Xing, Phys. Rev. D{\bf 61}, 073016 (2000), hep-ph/9909304;
E. Ma and G. Rajasekaran, Phys. Rev. D{\bf 64}, 113012 (2001), hep-ph/0106291;
P.~F.~Harrison and W.~G.~Scott, Phys.\ Lett.\  B{\bf 557}, 76 (2003) [arXiv:hep-ph/0302025]; 
S.-L. Chen, M. Frigerio, and E. Ma, Phys. Rev. D{\bf 70}, 073008 (2004) [Erratum-ibid. D{\bf 70}, 079905 (2004)] hep-ph/o404084;
F. Caravaglios and S. Morisi, hep-ph/0503234;
W. Grimus and L. Lavoura, JHEP {\bf 0508}, 013 (2005), hep-ph/0504153;
J. E. Kim and J. -C. Park, JHEP {\bf 0605}, 017 (2006), hep-ph/0512130;
R. N. Mohapatra, S. Nasri, and H. B. Yu, Phys. Lett. B{\bf 639}, 318 (2006), hep-ph/0605020;
R. Jora, S. Nasri, and J. Schechter, Int. J. Mod. Phys. A{\bf 21}, 5875 (2006), hep-ph/0605069;
M. Picariello, Int. J. Mod. Phys. A{\bf 23}, 4435 (2008), hep-ph/0611189;
Y. Koide, Eur. Phys. J. C{\bf 50}, 809 (2007), hep-ph/0612058;
A. Mondragon, M. Mondragon, and E. Peinado, Phys. Rev. D{\bf 76}, 076003 (2007), arXiv:0706.0354 (hep-ph);
A. Mondragon, M. Mondragon, and E. Peinado, AIP Conf. Proc. 1026: 164 (2008), arXiv:0712.2488 (hep-ph);
C.-Y. Chen and L. Wolfenstein, Phys. Rev. D{\bf 77}, 093009 (2008), arXiv:0709.3767.

\bibitem{GXZ}W-l. Guo, Z-z. Xing, and S. Zhou, Int.J.Mod.Phys. E{\bf 16}, 1 (2007).
\bibitem{SHY} S.-F. Ge, H.-J. He, and F.-R. Yin, arXiv:1001.0940.
\bibitem{DGH} S.-F. Ge, D.A. Dicus and H.-J. He, in preparation.
\bibitem{19}B.Pontecorvo, Sov.Phys.JETP {\bf 6}, 429 (1958); Z.Maki, M.Nakagawa, S.Sakata, Prog. Theor.Phys.{\bf 28}, 870 (1962).
\bibitem{BM} A. Barroso and J. Maalampi, Phys. Lett. B{\bf 132}, 355 (1983).
\bibitem{FLMPR}G.L.Fogli, E. Lisi, A. Marrone, A. Palazzo, A. M. Rotunno, 
Phys. Rev. Lett. {\bf 101}, 141801 (2008) [arXiv:0806.2649] and arXiv:0809.2936[hep-ph].
\bibitem{Fetal} G.L. Fogli {\em et. al.}, arXiv:0805.2517v3 [hep-ph]
and Phys. Rev. {\bf D78}, 033010 (2008).
\bibitem{MINOS}P. Adamson {\em et. al.}, Phys. Rev. Lett. {\bf 103}, 261802-1 (2009).
\bibitem{Lam} C.~S.~Lam, ``The Unique Horizontal Symmetry of Leptons'', Phys. Rev D {\bf 78}, 073015 (2008), [arXiv:0809.1185 [hep-ph]].
\end{thebibliography}
\end{document}